\documentclass[aps,prb,reprint,showpacs,floatfix, superscriptaddress]{revtex4-1}

\setcitestyle{square,numbers}
\usepackage[english]{babel}
\usepackage{amsmath,amsthm,amssymb}
\usepackage{amsfonts}
\usepackage{xspace}
\usepackage{bm}
\usepackage{graphicx}
\usepackage[separate-uncertainty = true, exponent-product = \times]{siunitx}
\usepackage{lmodern}
\usepackage{hyperref}
\usepackage{color}

\hypersetup{urlcolor=blue, colorlinks=true, citecolor=blue, linkcolor=blue}

\renewcommand{\dfrac}[2]{\frac{d#1}{d#2}}
\newcommand{\etal}{\emph{et~al.}}

\newcommand{\subref}[2][]{\ref{#2}\hyperref[#2]{#1}}

\begin{document}

\title{Spin Seebeck effect at microwave frequencies}


\author{Michael Schreier}
\email{michael.schreier@wmi.badw.de}
\affiliation{Physik-Department, Technische Universit\"at M\"unchen, Garching, Germany}

\author{Franz Kramer}
\affiliation{Physik-Department, Technische Universit\"at M\"unchen, Garching, Germany}

\author{Hans Huebl}
\affiliation{Nanosystems Initiative Munich, Munich, Germany}

\author{Stephan Gepr\"ags}
\noaffiliation

\author{Rudolf Gross}
\affiliation{Physik-Department, Technische Universit\"at M\"unchen, Garching, Germany}
\affiliation{Nanosystems Initiative Munich, Munich, Germany}

\author{Sebastian~T.~B. Goennenwein}
\affiliation{Nanosystems Initiative Munich, Munich, Germany}

\collaboration{\small{\emph{Walther-Mei\ss ner-Institut, Bayerische Akademie der Wissenschaften,
Garching, Germany}}}


\author{Timo Noack}
\author{Thomas Langner}
\author{Alexander~A. Serga}
\author{Burkard Hillebrands}

\author{Vitaliy~I. Vasyuchka}

\collaboration{\small{\emph{Fachbereich Physik und Landesforschungszentrum OPTIMAS, Technische Universit{\"a}t Kaiserslautern, Kaiserslautern, Germany}}}

\date{\today}

\begin{abstract}
We experimentally study the transient voltage response of yttrium iron garnet/platinum bilayer samples subject to periodic heating up to gigahertz frequencies. We observe an intrinsic cutoff frequency of the induced thermopower voltage, which characteristically depends on the thickness of the yttrium iron garnet film. The cutoff frequency reaches values of up to $\SI{350}{\mega Hz}$ in a $\SI{50}{\nano m}$ thick yttrium iron garnet film, but drops to below $\SI{1}{\mega Hz}$ for several-micrometer-thick films. These data corroborate the notion that the magnon spectrum responsible for the spin current emission in the spin Seebeck effect can be shaped by tuning the thickness of the ferromagnetic layer.
\end{abstract}
\maketitle

\section{Introduction}
Angular momentum currents have been proposed as an alternative to charge currents for the implementation of logic devices and effective magnetization control at the nanometer scale~\cite{Miron2011, Pai2012, Liu2012}. 
These so-called pure spin currents can be generated either by the interconversion of charge into spin currents by the spin Hall effect in heavy metals~\cite{Dyakonov1971a, Hirsch1999} or by stimulating magnetization dynamics in a ferromagnet (and other types of materials with long-range magnetic order) to serve as a spin battery in proximity to a spin sink. A collective stimulus to the magnetization can be achieved through resonant excitation (spin pumping~\cite{Tserkovnyak2002, Mosendz2010}) or nonresonantly by creation of a thermal nonequilibrium at the interface between the ferromagnet and the spin sink (spin Seebeck effect~\cite{Uchida2010}). While the spin Hall effect and spin pumping are reasonably well understood from a theoretical perspective, the detailed microscopic mechanism responsible for the spin Seebeck effect is still vividly discussed in the literature~\cite{Adachi2011, Adachi2013, Tikhonov2013, Xiao2010, Zhang2012, Hoffman2013, Chotorlishvili2013, Ritzmann2014, Etesami2015}. Experiments showed that the effect is nontrivially connected to the saturation~\cite{Uchida2014} and sublattice magnetization~\cite{Gepraegs2014} of the magnetically ordered material and that the effective temperature of the magnons, which are deemed responsible for the effect, is very close to that of the phonons~\cite{Agrawal2013}. Moreover, numerical studies~\cite{Schreier2013} and experiments~\cite{Flipse2014} suggested that the emitted spin current itself plays a significant role for the magnon thermalization process at the interface. The thermalization between any two systems in the solid state usually occurs on very short timescales (typically microseconds and smaller). Therefore the aforementioned experiments are of somewhat limited value for the clarification of the dynamic properties of the effect as all of them were performed on time scales corresponding to the static regime.\\
Two independent studies recently attempted to quantify the time scale for the magnon-phonon coupling relevant for the spin Seebeck effect by means of modulated laser heating, allowing for the generation of thermal gradients on a submicrosecond scale. Unfortunately the results of these experiments on yttrium iron garnet (YIG)/Pt samples were not conclusive. On the one hand, Agrawal \etal~\cite{Agrawal2014} observed a rolloff of the spin Seebeck voltage at frequencies below $\SI{1}{\mega Hz}$ in a several-micrometer-thick film and attributed it to a finite ``effective thermal magnon diffusion length.'' On the other hand, the experiments by Roschewsky \etal~\cite{Roschewsky2014} in much thinner YIG films suggest that the spin Seebeck effect should be robust even beyond several tens of megahertz.\\
We here resolve this apparent disagreement by performing systematic time-resolved spin Seebeck effect experiments with gigahertz experimental bandwidth on a series of samples with YIG thicknesses $\SI{50}{\nano m}\leq d_\mathrm{YIG}\leq\SI{53}{\micro m}$. We show that the characteristic decay time of the spin Seebeck effect changes as a function of the YIG layer thickness. The spin Seebeck voltages in our experiments show $\SI{3}{dB}$ cutoff frequencies as high as $\SI{0.35}{\giga Hz}$ in the thinnest investigated YIG films but drop to below $\SI{1}{\mega Hz}$ when the thickness of the YIG films exceeds $\SI{1}{\micro m}$. This is consistent with the notion that the energies of the magnons generating the measured spin Seebeck voltage increase with decreasing the YIG film thickness due to confinement~\cite{Kikkawa2015}. Furthermore, the evolution of the measured spin Seebeck voltage with the heating modulation frequency and the thickness of the YIG layer provides further evidence that the magnon spectrum stimulating the spin current emission characteristically changes with the YIG thickness. 
Our data complement recent results by Kehlberger \etal~\cite{Kehlberger2015} and Kikkawa \etal~\cite{Kikkawa2015}, who find a scaling of the absolute voltage levels with the YIG film thickness. They are further supported by theoretical predictions by Ritzmann \etal~\cite{Ritzmann2014} and Etesami \etal~\cite{Etesami2015}, who investigated magnon spectra subject to a thermal gradient. Providing the link between a large number of autonomous results reported in the literature, our experiments resolve crucial questions on the origin and characteristics of the spin Seebeck effect that were mostly a matter of speculations so far.\\

\section{Sample fabrication}
The YIG films used in our experiments were fabricated using two complementary methods allowing us to cover a large range of YIG film thicknesses. The thinner films ($d_\mathrm{YIG}\leq\SI{200}{\nano m}$) were grown in high vacuum 
using pulsed laser deposition (PLD). They were then covered with Pt films, without breaking the vacuum, by means of electron beam evaporation~\cite{Althammer2013}. Although the Pt layer thickness ($\approx\SI{10}{\nano m}$) varies by a few nanometers from sample to sample, this is not expected to have a significant impact on our analysis. Thicker YIG films are commercially fabricated by means of liquid phase epitaxy (LPE) 
and subsequently covered with Pt. In our analysis we assume the measured thermal voltages to originate exclusively from the spin Seebeck effect and a potential spurious contribution due to the anomalous Nernst effect. This is in agreement with the notion that the proximity polarization in Pt is very small~\cite{Geprags2012,*Geprags2013}, such that anomalous Nernst-type effects are at least two orders of magnitude smaller than the spin Seebeck effect in YIG/Pt~\cite{Kikkawa2013}.\\

\begin{figure}%
\includegraphics[width=85mm]{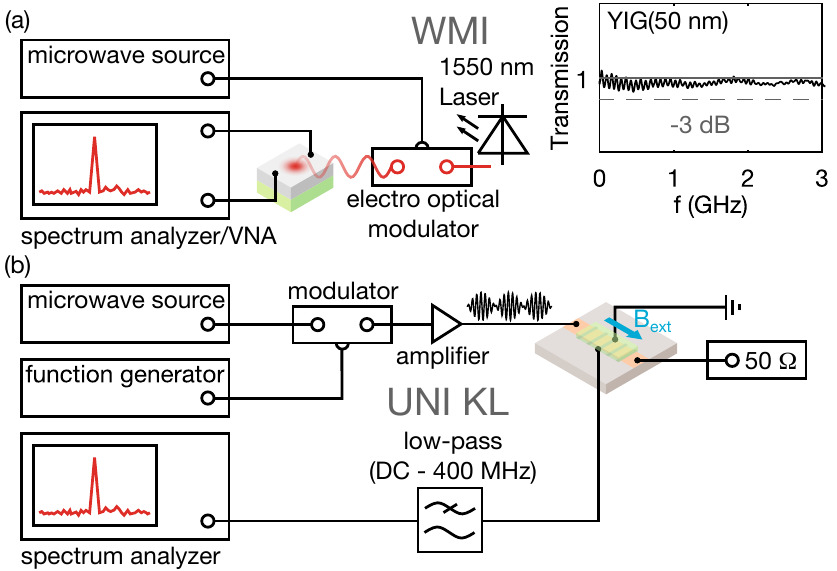}%
\caption{(a) At the WMI the thermal voltages generated by an intensity-modulated laser are recorded by a spectrum analyzer. The inset exemplarily shows the transmission $T=S_{21}/(1-S_{11})$ as obtained from the $S$-parameter measurements on the YIG($\SI{50}{\nano m}$) sample. (b) At the UNI KL samples are placed on a microstrip line which is heated periodically by high-power microwave pulses. After filtering out the original microwave frequency, voltages are detected at the pulse frequency.}
\label{fig:setup}%
\end{figure}

\section{Measurement setup}
Two different experimental approaches were employed to investigate the spin Seebeck effect dynamics. At the Walther-Mei{\ss}ner-Institut (WMI) a continuous-wave solid-state laser with a wavelength of $\SI{1550}{\nano m}$ is modulated at a frequency $f$ by an electro-optical modulator (EOM) to generate a sinusoidal intensity modulation of the laser beam with a peak-to-peak amplitude of about $\SI{30}{\milli W}$ and thus generates a time-varying thermal gradient in the samples. The time-varying spin Seebeck voltage is detected and averaged by a spectrum analyzer (Fig.~\subref[a]{fig:setup}). In our setup we ensured via $S$-parameter measurements that bandwidth limitations of the electric circuit are not at the origin of the observed frequency dependence.\\
At the Technische Universit{\"a}t Kaiserslautern (UNI KL) samples are placed on top of a microstrip line, covered by an insulating layer. By application of a $\SI{30}{dBm}$ sine-modulated microwave ($f_\mathrm{mw}=\SI{6.875}{\giga Hz}$) current to the microstrip line, eddy currents are induced in the Pt, which, in turn, generate a time-varying thermal gradient across the YIG/Pt interface~\cite{Agrawal2014a}. After passing a low-pass filter, the voltages are detected by a spectrum analyzer at the microwave modulation frequency $f$ (Fig.~\subref[b]{fig:setup}). The induced temperature variation is of the order of a few tens of kelvins~\cite{Schreier2013,Agrawal2014} for both experimental approaches.\\
Both experiments, upon repeating the procedure for different modulation frequencies, directly yield the characteristic frequency response transfer function $|V_\mathrm{SSE}|(f)$, i.e., the magnitude of the voltage response as a function of the frequency of the applied temperature gradient. The consistent results on samples of comparable thickness demonstrate the compatibility of both approaches. With applied magnetic fields of the order of $\SI{25}{\milli T}$ the experiments are performed far below ferromagnetic resonance and parametric excitation conditions.\\

\section{Experimental results}
\begin{figure}%
\includegraphics[width=\columnwidth]{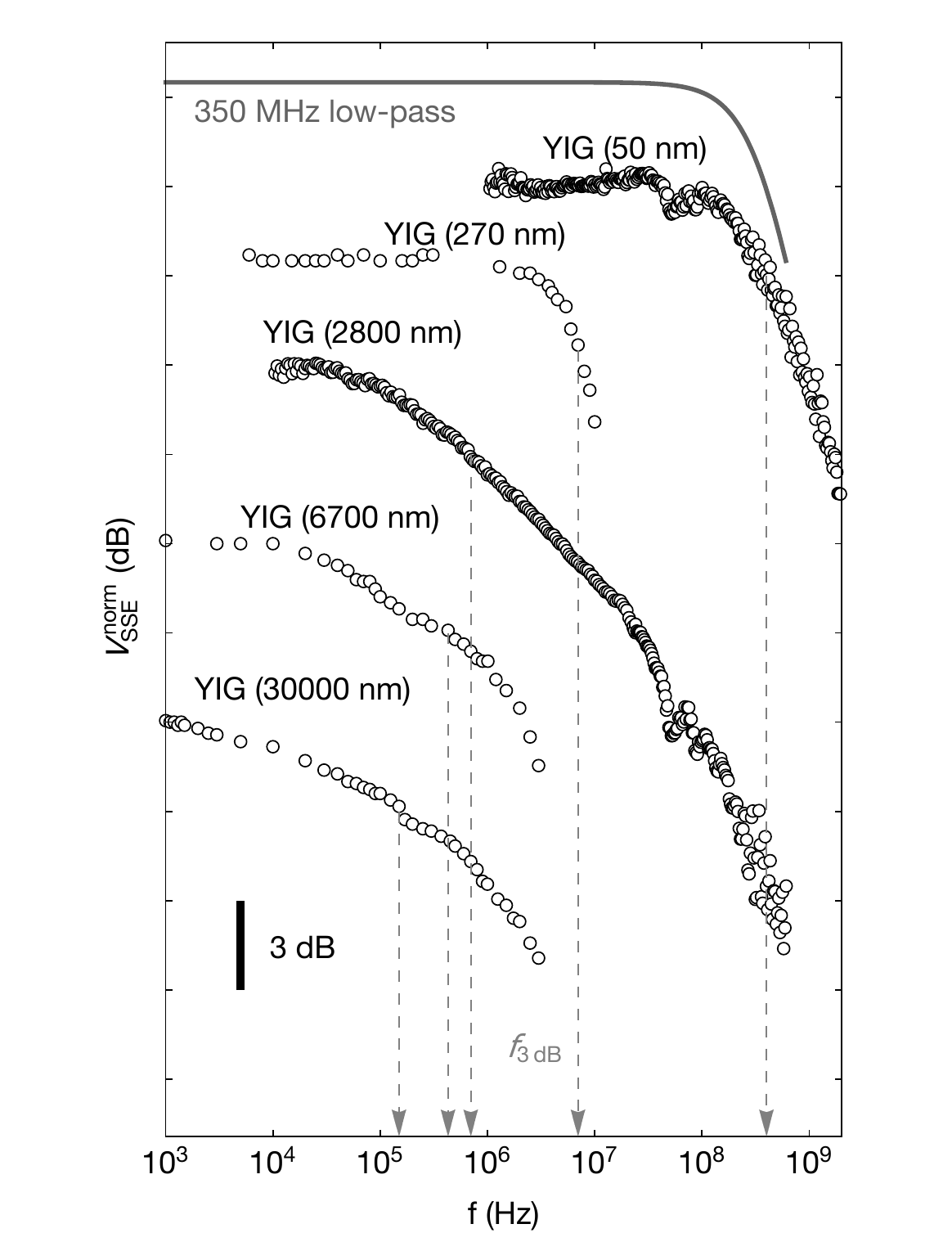}\\%
\caption{Normalized spin Seebeck voltage $V_\mathrm{SSE}(f)/V_\mathrm{SSE}(0)$ as a function of the modulation frequency of the applied temperature gradient for the $\SI{50}{\nano m}$ and $\SI{2800}{\nano m}$ thick YIG films recorded at the WMI and the $\SI{270}{\nano m}$, YIG($\SI{6700}{\nano m}$) and $\SI{30000}{\nano m}$ thick YIG films recorded at the UNI KL. An offset between the data for different samples has been added for clarity. The dashed lines indicate the $\SI{3}{dB}$ points and the solid line depicts the behavior of a $\SI{350}{\mega Hz}$ low-pass filter for comparison.}%
\label{fig:transfer}%
\end{figure}

\begin{figure}%
\includegraphics[width=85mm]{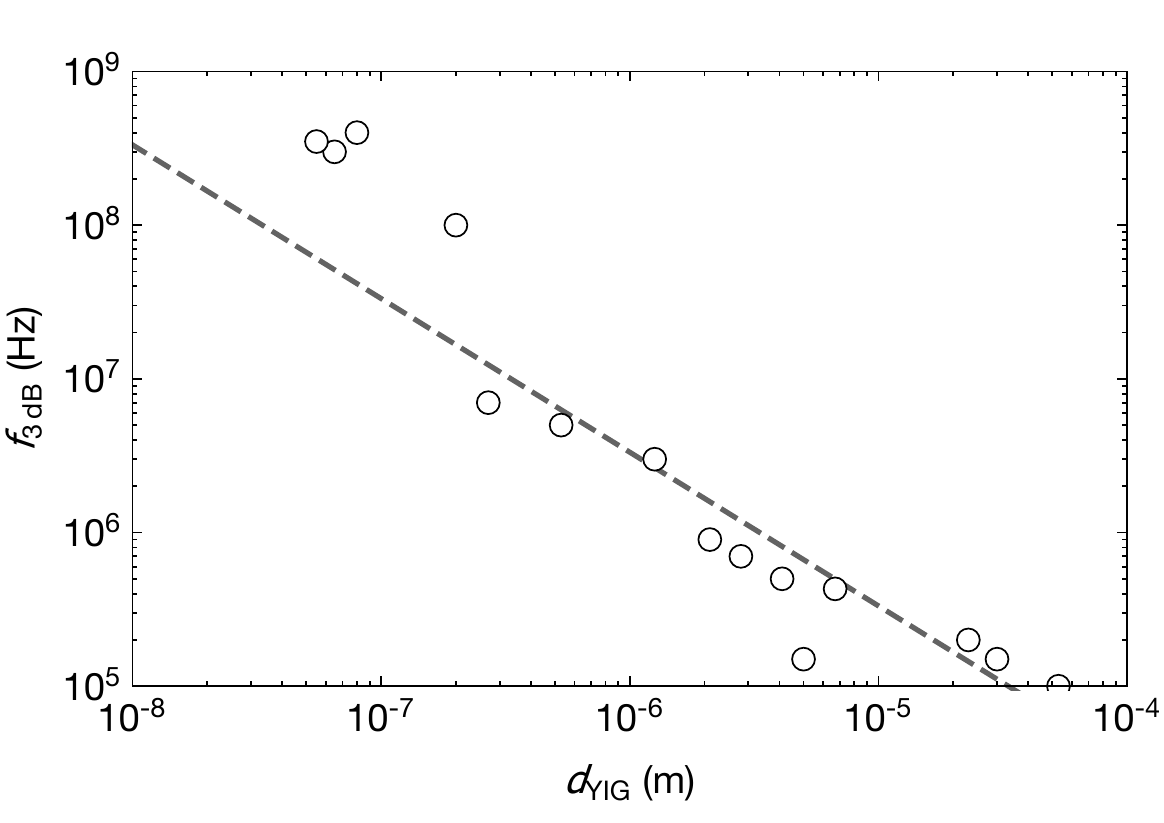}%
\caption{The $\SI{3}{dB}$ rolloff frequency as a function of the YIG film thickness. $f_{\SI{3}{dB}}$ gradually decreases as approximately $d_\mathrm{YIG}^{-1}$ indicated by the dashed line.}%
\label{fig:fvsd}%
\end{figure}

The transfer functions $|V_\mathrm{SSE}|(f)$ for samples YIG($\SI{50}{\nano m}$), YIG($\SI{270}{\nano m}$), YIG($\SI{2800}{\nano m}$), YIG($\SI{6700}{\nano m}$), and YIG($\SI{30000}{\nano m}$) are exemplarily shown in Fig.~\ref{fig:transfer}. The absolute voltage levels follow the trend reported in Refs.~\onlinecite{Kehlberger2015, Kikkawa2015}, starting from about $\SI{100}{\nano V}$ in the YIG($\SI{50}{\nano m}$) sample to several microvolts in micrometer thick films, where small sample-to-sample variations eventually cloud any further scaling. However, here the data have been normalized to the respective dc value for each sample for clarity. Evidently, the spin Seebeck voltage on the $\SI{50}{\nano m}$ film remains at its dc level up to much higher frequencies as compared to the other samples. Moreover, the voltage response for frequencies above the $\SI{3}{dB}$ point is markedly different for the different samples. For the $\SI{50}{\nano m}$ and $\SI{270}{\nano m}$ films the transfer function resembles a classical first order low-pass, decaying as $1/f$ for $f\gg f_{\SI{3}{dB}}$. In contrast, $|V_\mathrm{SSE}|(f)$ measured for the $2800$, $6700$, and $\SI{30000}{\nano m}$ film decays at a much lower rate. A gradual change of both the cutoff frequency as well as the shape of the transfer function is observed as a function of YIG thickness in our sample series (not shown here).\\
Since the transfer functions of the investigated samples clearly cannot all be modeled with the same approach (e.g., a low-pass filter as above), we determine the cutoff (\SI{3}{dB}) frequencies $f_{\SI{3}{dB}}$ for all samples solely by the frequency at which $V_\mathrm{SSE}(f_{\SI{3}{dB}})=V_\mathrm{SSE}(f\to0)/\sqrt{2}$. The result is plotted in Fig.~\ref{fig:fvsd} and indicates a power-law like behavior of the cutoff frequency $f_{\SI{3}{dB}}\propto d^\beta$ as a function of the YIG film thickness $d$ with an exponent $\beta\simeq-1$. The larger $f_{\SI{3}{dB}}$ values for the thinnest films can be attributed to the more pronounced magnetization damping [cf. Eq.~\eqref{eq:Etesami}] in samples grown by pulsed laser deposition compared to films fabricated via liquid phase epitaxy. Lacking unequivocal evidence of a plateaulike behavior at low frequencies it is possible that the \emph{true} $f_{\SI{3}{dB}}$ value for our thickest ($>\SI{10}{\micro m}$) films is lower than what we extract here. Considering, however, that the extracted values smoothly extend the trend established by the thinner samples, the potential error is unlikely to be excessive.\\

\section{Discussion}
Within the spin mixing interface conductance formalism~\cite{Brataas2000} an energy imbalance between the two sides of a ferromagnet/normal metal interface~\cite{Weiler2013} leads to the flow of angular momentum across the interface. For the spin Seebeck effect this energy difference may be formulated in terms of a temperature difference between electrons in the normal metal and magnons in the ferromagnet. Such a temperature difference may originate, e.g., from diffusive~\cite{Xiao2010} or stochastic~\cite{Ritzmann2014} transport or by considering ``subthermal'' magnons~\cite{Tikhonov2013, Hoffman2013, Boona2014}. While data at elevated temperatures are sparse, low-temperature measurements~\cite{Boona2014} and the relatively long lifetime of the small-wave-number $k$ magnons~\cite{Spencer1962} even at room temperature suggest that the thermalization process between magnons and phonons is the limiting factor for high-frequency spin Seebeck excitations. The interaction with the other thermal reservoir, the electrons in the normal metal, can be assumed instantaneous in the experimentally accessed frequency range as spin current emission has been demonstrated for much higher frequencies, e.g., in spin pumping~\cite{Mosendz2010} or spin Hall magnetoresistance~\cite{Lotze2014} experiments.\\
To explain the observed frequency dependence of the spin Seebeck voltage we start from a steady-state model~\cite{Hoffman2013, Schreier2013, Flipse2014}. The energy supplied to the Pt layer by the laser (or microwave) heating raises the temperature of the Pt layer (electrons and phonons are approximately in thermal equilibrium due to the fast electron-phonon scattering). Energy and angular momentum are then transferred by the spin transfer torque from the electrons in the Pt to the magnons in the YIG, raising the magnon temperature $T_\mathrm{m}$. Due to magnon-phonon scattering the energy is finally transferred to the phonon system and the heat sink. In the stationary situation a steady-state spin current is generated across the YIG/Pt interface which is proportional to the temperature difference $\Delta T = T_\mathrm{e}-T_\mathrm{m}$ of the electrons in Pt and the magnons in YIG. It is obvious that a finite $\Delta T$ and, hence, a finite spin Seebeck voltage is obtained only if the magnons couple to the phonons. Otherwise, $\Delta T\rightarrow 0$. This is important for the nonstationary case. A finite $\Delta T$ is obtained only on time scales longer than the characteristic magnon-phonon interaction time $\tau_\mathrm{mp}$ in YIG. On shorter time scales, the energy cannot be transferred from the magnon to the phonon systems, resulting in $\Delta T\rightarrow 0$. In a simple relaxation time approach, the time evolution of the temperature difference $\Delta T_\mathrm{mp}$ between the magnons and phonons in YIG can be expressed as~\cite{Sanders1977}
\begin{equation}
	\dfrac{}{t}\Delta T_\mathrm{mp}=-\frac{\Delta T_\mathrm{mp}}{\tau_\mathrm{mp}}.
\label{eq:DTmp}
\end{equation}
This is trivially solved by $\Delta T_\mathrm{mp}(t)\propto e^{-\frac{t}{\tau_\mathrm{mp}}}$, which transforms to
\begin{equation}
	\left|\Delta T_\mathrm{mp}(\omega)\right|\propto \frac{1}{\sqrt{1+\left(\omega\tau_\mathrm{mp}\right)^2}}
\label{eq:DTmpSolw}
\end{equation}
in the frequency domain. The transfer function derived from this model indeed describes the data of the $50$ and $\SI{270}{\nano m}$ films reasonably well. Following this reasoning the cutoff frequency in the $\SI{50}{\nano m}$ film thus corresponds to a characteristic interaction time between magnons and phonons of $\tau_\mathrm{mp}=1/(2\pi f_{\SI{3}{dB}})\approx\SI{450}{\pico s}$. This is consistent with estimates in the literature~\cite{Schreier2013,Rezende2014} which put $\tau_\mathrm{mp}$ at a few hundred picoseconds for high-energy magnons. We note that in the simple relaxation time approach with a single, frequency independent relaxation time the extracted $\tau_\mathrm{mp}$ likely is a weighted average over the entire magnon spectrum contributing to the emission of the measured spin current. Keeping in mind that $\tau_\mathrm{mp}$ rather is frequency dependent further explains why the transfer function for the thicker films deviates significantly from Eq.~\eqref{eq:DTmpSolw}. In thermal equilibrium and in the absence of any spatial variation of the temperature, the magnon population is distributed according to Bose-Einstein statistics and can be described with a single magnon temperature $T_\mathrm{m}$. A thermal gradient, however, is accompanied by a flow of magnons along the same direction, with the diffusion length of individual magnons decreasing with their energy~\cite{Schreier2013}. The different diffusion lengths become important at the edges (interfaces) of the magnetic layer at which only magnons closer than their respective diffusion length can accumulate. Increasing the magnetic layer thickness then favors lower energetic magnons with their longer diffusion lengths in the interface accumulation. This also implies that increasing the thickness of the ferromagnet leads to a growing deviation from Bose-Einstein statistics right at the interface and the concept of a single temperature is no longer well defined~\cite{Casas-Vazquez2003}. According to numerical simulations~\cite{Ritzmann2014} this effect is small for our thinnest films but should be more significant in thicker ones. At any rate, the magnon spectrum is expected to show an exponential relaxation, similarly to Eq.~\eqref{eq:DTmp}. The lowest energetic magnons (i.e. those with frequencies close to the ferromagnetic resonance frequency) feature interaction times with phonons exceeding a microsecond and one would indeed expect the effective cutoff frequency to shift to lower values in thicker YIG films. More specifically, Etesami \etal~\cite{Etesami2015} derived the functional dependence of the $\SI{3}{dB}$ rolloff frequency $f_{\SI{3}{dB}}$ for a given magnon mode $n\in\mathbb{N}_0$ on $d_\mathrm{YIG}$ as 
\begin{equation}
	f_{\SI{3}{dB}}^{n}=\alpha\frac{\gamma}{2\pi}\left[B_0+\frac{2A}{M_\mathrm{s}}\left(\frac{n\pi}{d_\mathrm{YIG}}\right)^2\right],
\label{eq:Etesami}
\end{equation}
where $\alpha$ is the magnetization damping, $\gamma$ is the gyromagnetic ratio, $B_0$ is the external magnetic field, $A$ is the exchange constant, and $M_\mathrm{s}$ is the saturation magnetization. The full transfer function is then a linear combination of the low-pass behavior of the individual magnon modes. As $d_\mathrm{YIG}$ increases, the mode number $n$ for magnons still effective at a given energy increases and more lower energetic modes become available. Since these lower energetic modes feature smaller $f_{\SI{3}{dB}}^n$ values [cf. Eq.~\eqref{eq:Etesami}] the total $f_{\SI{3}{dB}}$ shifts downwards, at a rate determined by the mode occupation number, i.e. the composition of the magnon spectrum. The $d_\mathrm{YIG}^{-1}$ behavior observed in Fig.~\ref{fig:fvsd} is qualitatively consistent with an increased weight of the lower energetic end of the magnon spectrum in thicker films. In any case, the $f_{\SI{3}{dB}}$ values plotted in Fig.~\ref{fig:fvsd} may be understood as approximate values of the dominant magnon mode, which we emphasize becomes increasingly crude when many modes with different $f_{\SI{3}{dB}}^{n}$ values contribute similarly to the total signal in thicker YIG films. With this caveat in mind $f_{\SI{3}{dB}}$ in Fig.~\ref{fig:fvsd} is related to the characteristic magnon energy stimulating the spin current emission via
\begin{equation}
	E=\hbar\omega=\frac{2\pi\hbar}{\alpha}f_{\SI{3}{dB}}.
\label{eq:Energy}
\end{equation}
Considering the different damping characteristics of the films grown by pulsed laser deposition ($\alpha\approx10^{-3}$~\cite{Althammer2012}) and liquid phase epitaxy ($\alpha\approx10^{-4}$~\cite{Heinrich2011, Hung2010, Kurebayashi2011, Hoffman2013}), we obtain characteristic energies of the order of $\SI{1}{\milli eV}$ in the thinnest films. For the $\SI{50}{\nano m}$ YIG film specifically Eq.~\eqref{eq:Etesami} suggests the $n=7,\ k\approx\SI{4e8}{m^{-1}}$ mode as the dominant magnon mode ($n=17,\ k\approx\SI{2e8}{m^{-1}}$ for the LPE grown $\SI{270}{\nano m}$ YIG film). 
Converting the characteristic energy of this particular sample into an \emph{effective} temperature we obtain $T_\mathrm{eff}\approx\SI{17}{K}$. This is in good agreement with the results by Boona and Heremans~\cite{Boona2014} and Jin \etal~\cite{Jin2015} who give an upper temperature limit for the magnons contributing to the spin current of $30-\SI{40}{K}$. We note, however, that in the context of measurements at room temperature it is highly ambiguous to refer to a subset of the magnon spectrum by means of its ``temperature'', rather than its energy. From a different perspective, although themed ``low-energy magnons'' here, the characteristic frequencies of the magnons (up to a few terahertz) are substantially larger than those typically investigated in ferromagnetic resonance experiments (a few gigahertz). Still, the characteristic energies of the order of a $\SI{}{meV}$ at best render them rather low in energy compared to the energy scale suggested by the room-temperature measurements.\\
Finally, it is important to note that some caution is required regarding the specificity of the parameters derived from our experiments. It is clear that there is some leeway in the interpretation of the raw experimental data and that the different sample fabrication methods and related variations in, e.g., magnetic damping in the films affect the accuracy of the derived numerical values. Additionally the possible influence of a thermal skin effect~\cite{Agrawal2014} is not explicitly accounted for in the above analysis. While a skin effect will influence the magnitude and phase of the temperature modulation depending on the distance to the interface, it leaves the modulation frequency unaffected. 
Simultaneously the magnitude of the attenuation at each point should be agnostic to the YIG film thickness and thus fundamentally the same for all samples. In YIG films thinner than the magnon mean free path~\cite{Agrawal2014}, however, some of the observed attenuation could be a residue of the above phenomenon. Nevertheless, the extracted values appear to be consistent with the literature and are further supported by the $f_{\SI{3}{dB}}(\SI{500}{\nano m})$ value of some $\SI{10}{\mega Hz}$ inferred from numerical simulations in Ref.~\cite{Etesami2015}.\\

\section{Conclusion}
In summary, we investigated the frequency dependence of the spin Seebeck effect and identified characteristic response times which depend on the thickness of the YIG layer. Specifically we find that the characteristic response time scales approximately as the inverse thickness of the films. This is consistent with recent experimental and theoretical studies and supports the notion that high-frequency magnons dominate the effect in thin films but are less important for its absolute magnitude and frequency dependence in thicker ones. Our results allow us to develop a consistent picture of the physics behind the spin Seebeck effect, providing a link between several recent experimental~\cite{Kehlberger2015,Kikkawa2015,Boona2014,Roschewsky2014,Agrawal2014,Jin2015} and theoretical~\cite{Ritzmann2014,Etesami2015,Rezende2014} results. They could further be relevant for the numerical evaluation of the spin Seebeck effect~\cite{Schreier2013,Rezende2014} and related~\cite{Flipse2014} phenomena. Going beyond the scope of the experiments presented here, further insights might be gained by systematically studying the effect of large magnetic fields and low temperatures on the dynamics. Large magnetic fields have already been demonstrated to affect the dc properties~\cite{Kikkawa2015}, an effect which should also translate to the transient response. The approach presented here could also provide further insight into the spectral composition of the thermally induced spin currents in compensated garnets~\cite{Gepraegs2014} with nontrivial contributions from different magnetic sublattices.\\

We thank Michaela Lammel for the fabrication of YIG/Pt samples at the WMI and Moritz Geilen, Bj\"orn Heinz, and the Nano Structuring Center for the sample preparation at the UNI KL. Financial support from the DFG via SPP 1538 ``Spin Caloric Transport,'' Projects No. GO 944/4-2 and No. SE 1771/4-2, is gratefully acknowledged.

%

\end{document}